\documentclass[twocolumn,aps]{revtex4}

\begin{document}

\title{Minimal Supersymmetric Pati-Salam Theory: 
Determination of Physical Scales.}

\author{Alejandra Melfo$^{(1,2)}$ and 
Goran Senjanovi\'c$^{(1)}$}
\affiliation{$^{(1)}${\it International Centre for Theoretical Physics,
34100 Trieste, Italy }}

\affiliation{$^{(2)}$ {\it Centro de Astrof\'{\i}sica Te\'orica, Universidad de
Los Andes, M\'erida, Venezuela }}

\begin{abstract}
We systematically study the minimal supersymmetric
Pati-Salam theory, paying special  
attention to the unification constraints.  We find that the $SU(4)_c$
scale $M_c$ and the Left-Right 
 scale $M_R$ lie in the range $10^{10} GeV<M_c < 10^{14} GeV \, , 
\; 10^{3} GeV < M_R <10^{10}  
GeV$ (with single-step breaking at $10^{10} GeV$), giving a
potentially accessible scale of parity breaking.  
The theory includes the possibility of having doubly-charged supermultiplets at
the supersymmetry breaking scale;  
 color octet states with mass $\sim M_R^2/M_c$;
magnetic monopoles of intermediate mass  
that do not conflict with cosmology, and a ``clean'' (type I) form for the
see-saw mechanism of neutrino mass.  

\end{abstract}
\maketitle

\section{Introduction}

The theories of Grand Unification owe their origin to the beautiful
idea of quark-lepton symmetry put forward  
about thirty years ago by Pati and Salam \cite{ps74}. The minimal
and original realization is based on a 
$G_{PS} = SU(2)_L\times SU(2)_R\times SU(4)_c$ symmetry, and has been
studied thoroughly in the  
past (for recent work see for example \cite{jkls00,bm01}
 and references therein). It is characterized by a number of interesting and
important features 

\begin{enumerate}
\item it incorporates Left-Right (LR) symmetry \cite{ps74,mp75,sm75} 
(on top of the
quark-lepton symmetry mentioned above), which leads naturally to the
spontaneous breaking of parity and charge conjugation \cite{sm75,s79}.
 
\item incorporates a see-saw mechanism for small neutrino masses
\cite{grs79,y79,ms80a,glashow79}

\item predicts the existence of magnetic monopoles \cite{p74,t74,s85}

\item leads to rare processes such as $K_L \to \mu \bar e$ through the
lepto-quark gauge bosons (with however a negligible rate for $M_{PS}
\geq 10 GeV$) \cite{ps74}

\item in the case of single-step breaking, predicts the scale of
quark-lepton (and Left-Right) unification \cite{pss83}
 
\item allows naturally for $\Delta B =2$ process of $n-\bar n$ oscillations
(with however a negligible rate unless there are light diquarks in the
$TeV$ mass region)\cite{k70,g79,mm80,cm99}.

\item last but not least, it allows for implementation of the
leptogenesis scenario, as suggested by 
 the see-saw mechanism \cite{fy86}. 

\end{enumerate}

Since the scale of unification is large (see below), the Pati-Salam
(PS) model
can turn out to be indistinguishable from a less unifying theory, such
as LR, or a more unifying one such as SO(10). The issue we want to address
here is, what could be the smoking gun of a Pati-Salam model?

We believe that this issue should be investigated in the context of
the minimal PS theory. Of course,  
 one wishes to
appeal to supersymmetry as the usual  
protection mechanism for large hierarchies. Now, the supersymmetric
standard model includes  potentially catastrophic $d=4$ operators
leading to proton decay. The supersymmetric PS
model, as any model of the renormalizable see-saw mechanism, solves
this problem since it leads to R-parity being an exact symmetry to all
orders in perturbation theory \cite{amrs99}.  As a bonus,
 the LSP is stable and becomes a natural candidate for the dark matter
of the universe. 

We will therefore systematically study the minimal supersymmetric
Pati-Salam theory, paying  
special attention to possible low energy predictions. 
We allow for two-step breaking
\begin{eqnarray}
SU(2)_L\times SU(2)_R\times SU(4)_c  \nonumber \\ 
\longrightarrow SU(2)_L\times SU(2)_R\times U(1)_{B-L}SU(3)_c \nonumber\\
 \longrightarrow SU(2)_L\times U(1)_Y\times SU(3)_c 
\label{breaking}
\end{eqnarray}
Our findings are as follows. If one ignores non-renormalizable terms,
the $M_c$ scale is in the range 
\begin{equation}
10^{10} GeV < M_c < 10 ^ {14} GeV 
\end{equation}
while at the same time
\begin{equation}
10^{10} GeV >M_R > 10 ^ {3} GeV 
\end{equation}

In other words, lower $M_R$ implies larger $M_c$. We find it
remarkable that unification allows $M_R$ to be   
close to the experimental limit, and thus potentially observable. 

Of course, the lower $M_R$, the more fine-tuning needed in order to
have neutrinos light. But still, strictly  
speaking, Dirac neutrino mass can be arbitrary, and thus neutrinos
could be light. At the same time $M_c$ is pushed to its upper limit
$\sim 10^{14} GeV$ which leads to the well-known monopole problem of
theories with symmetry breaking scales above $10^{11} GeV$. One can 
invoke inflation (or some other mechanism, \cite{lp80,dms95,dlv98})
 to solve the  problem, but
this simply renders the $SU(4)_c$ symmetry invisible.  

Another extreme is the single-step breaking with $M_R \sim M_c \simeq
10^{10} GeV$. In this case monopoles have 
a mass $m_M \simeq 10^{11} GeV$, and they are perfectly compatible with
all the experimental, astrophysical and cosmological data. At the same
time one has the usual leptogenesis scenario \cite{b02} and
naturally small neutrino masses; and  furthermore $SU(4)_c$ could be
observable through the possible discovery of magnetic monopoles.  

Another interesting consequence of this theory is a ``clean''(type I) 
see-saw mechanism, as we discuss later.

Most important, this theory includes the possibility of  existence 
 of light doubly charged supermultiplets (as noticed in the
past  \cite{ams97,cm97})
with masses in the $TeV$ region, and a potentially light
 (for $M_R \ll M_c$) color octet supermultiplet with 
mass $\sim M_R^2/M_c$, which could also be as low as $TeV$.
Relatively light color octets are known to be present in other theories.
 Even in the minimal supersymmetric $SU(5)$ they could lie much below the 
GUT scale \cite{bfy96}.

Non-renormalizable effects suppressed by $M_{Pl}$ become important if
$M_R^2 > M_W M_{Pl}$. In this case, the single-step breaking at
 $M_R \sim M_c \simeq
10^{10} GeV$ turns out to be  the only possibility, and the doubly
charged supermultiplets become potentially observable with masses in
the $TeV$ range. However, the color octet becomes heavy, making the
theory indistinguishable from the LR model. 

A more likely possibility is the existence of
non-renormalizable terms cut-off by a much lower scale than $M_{Pl}$. 
Namely, the theory is not asymptotically free above $M_c$, and the
gauge coupling becomes strong at $M_F\simeq 10 M_c$. Including $1/M_F$
terms (with $M_R^2 > M_W M_F$)
leads to $M_R\geq 10^7 GeV$, $M_c \geq 10^{12} GeV$, and the color
octet and doubly-charged multiplets have the same mass, $\sim
M_R^2/M_c$. Again, these particles could be observable and reveal four
colour unification. 
In what follows we  discuss these findings at length.

\section{The model: minimal PS theory with two-step breaking}

It is easy to see that  (\ref{breaking}) can be achieved
 with the following minimal set of Higgs-like
supermultiplets:
(the numbers in parenthesis indicate the $G_{PS}$ representations)
\begin{eqnarray}
&A (1,1,15)&\nonumber \\
&\Sigma  (3,1,10) , 
\;\bar\Sigma  (3,1,\bar{10}) , \;\Sigma_c (1,3,\bar{10}) , 
\;\bar\Sigma_c (1,3,10) &
\label{asigma}
\end{eqnarray}

The matter supermultiplets are
\begin{equation}
\psi (2,1,4) , \; \psi_c (1,2,\bar 4)
\end{equation}
and the minimal light Higgs multiplet is
\begin{equation}
\phi (2,2,1)
\end{equation}

The most general superpotential for the fields (\ref{asigma}) is
\begin{equation}
W = m Tr A^2 + M Tr( \Sigma\bar\Sigma + \Sigma_c\bar\Sigma_c) +  Tr(
\Sigma A \bar\Sigma - \Sigma_c A \bar\Sigma_c )
\end{equation}
where we assume the following transformation properties under Parity
\begin{equation}
\Sigma \to \Sigma_c , \quad \bar\Sigma \to \bar \Sigma_c , \quad A \to -A
\end{equation}
We choose $A$ to be a parity-odd field in order to avoid flat
directions connecting Left- and Right-breaking minima.

Notice that under $SU(3)_c$, the symmetric representation 10 of
$SU(4)_c$ is  decomposed  as 10 = 6 + 3 + 1. Clearly, only the singlets of
$SU(3)_c$ in the $\Sigma$ fields, which we shall denote as $\Delta,
\bar\Delta, \Delta_c$ and $\bar\Delta_c$, can take non-vanishing vevs.

This allows for the supersymmetry-preserving symmetry breaking pattern
\begin{equation}
<A> = M_c \, {\rm diag} (1,1,1,-3)
\label{avev}
\end{equation}
\begin{eqnarray}
<\Delta> &=&<\bar\Delta> = 0 \label{sigmaleft} \\
<\Delta_c>= M_R \left(\begin{array}{cc}
0 & 1 \\
0 & 0 \end{array} \right)   &,&
<\bar\Delta_c> =  M_R \left(\begin{array}{cc}
0 & 0 \\
1 & 0 \end{array} \right) \label{sigmavev}
\end{eqnarray}
(where the matrix in (\ref{avev}) is in $SU(4)$ space and those of 
(\ref{sigmavev})  are in $SU(2)_R$ space),
with
\begin{equation}
M_c \simeq M , \quad M_R \simeq \sqrt{M m}
\end{equation}

The mass spectrum is easy to compute
\begin{center}
\begin{tabular}{l|l}
$M_c$ &\hspace{0.3cm}all states in $\Sigma,\bar\Sigma$;\\
 & \hspace{0.3cm} the states in $A$ except for an octet of color \\ 
& \hspace{0.3cm} the states in $\Sigma_c, \bar\Sigma_c$ except for $\Delta_c $ and
$\bar\Delta_c $  \\ 
& \\
$M_R$ &\hspace{0.3cm} the fields $\Delta_c(1,3,1) $ and
$\bar\Delta_c(1,3,1) $ of 
$\Sigma_c, \bar\Sigma_c$,\\
  &\hspace{0.3cm} except for the components
$\delta_c^{++},\bar\delta_c^{++}$\\ &\hspace{0.3cm} and a 
combination of the singlet components \\
 & \\
$M_R^2/M_c$ &\hspace{0.3cm} color octet in A\\
 & \\
$\Lambda_{SUSY}$ &\hspace{0.3cm} supermultiplets
$\delta_c^{++},\bar\delta_c^{++} $ \\ &\hspace{0.3cm}  and 
a combination of the singlets in  $\Delta_c $ and
$\bar\Delta_c $ \\
  &\hspace{0.3cm} (and all the MSSM superpartners) \\
 & \\
\end{tabular}
 \end{center}
\vspace{0.5cm}

Some comments are in order. The color octet mass is clear, $m
\sim M_R^2/M_c$, but the situation with
$\delta_c^{++},\bar\delta_c^{++} $ is more subtle, and although it has
been discussed before \cite{cm97} it is worth repeating here. The fields
$\Delta_c(1,3,1) $ and $\bar\Delta_c(1,3,1) $ responsible for the
scale $M_R$ are coupled to the $(1,1,1)$ field in $A$ and so appear in
the superpotential only through terms with
 $Tr \Delta_c\bar\Delta_c $. This implies a larger, accidental $SU(3)$
 symmetry broken down to $SU(2)$, hence five Nambu-Goldstone bosons. But
the gauge symmetry $SU(2)_R\times U(1)_{B-L}$ is broken down to
$U(1)_Y$, so that three of them are eaten, leaving us with massless
states $\delta_c^{++},\bar\delta_c^{++}$. Of course, they become
massive when supersymmetry gets broken, and so they have masses of
order $TeV$. This is the most interesting prediction of SUSY LR
 (PS) theories.

However, there is a problem associated with this. Namely, the
charge-preserving vacuum lies on a flat direction \cite{km93}
\begin{equation}
<\Delta_c> = M_R \left(\begin{array}{cc}
0 & \cos\theta \\
\sin\theta & 0 \end{array} \right) \quad
<\bar\Delta_c> = M_R \left(\begin{array}{cc}
0 & \sin\theta \\
\cos\theta & 0
 \end{array} \right)
\end{equation}
(in other words, the potential is $\theta$-independent, and charge
conservation requires $\theta=0$ or $\theta = \pi/2$).
If the soft breaking terms break also $SU(2)_R$, then, of course, one
may eliminate the charge breaking vacua. If not, one
would have to appeal to higher dimensional operators (which in general break the accidental
$SU(3)$ symmetry).  We discuss this possibility in Section IV.

On top of the above states are the usual quarks and leptons in 
\begin{equation}
\psi(2,1,4) ; \quad \psi_c(1,2,\bar 4)
\end{equation}
and the two MSSM Higgs doublet superfields belonging to a combination 
of the (2,2,1) and the (2,2,15) fields (see below). The fermions
$\psi$ and $\psi_c$ provide a complete representation (we assume no new
matter states) and can be used to normalize the $U(1)$ generators,
such as $Y$ or $B-L$.

\section{Unification constraints} 

From the charge formula in LR
symmetric theories, 
we have $
Y/2= I_{3R} + (B-L)/2$. 
Using the fact that $(B-L)/2$ is a generator of $SU(4)_c$, we can
properly normalize the generators. It is then easy to see that the
combination 
\begin{equation}
\Delta\alpha^{-1} \equiv \alpha_Y^{-1} - \alpha_L^{-1} -
 \frac{2}{3}\alpha_c^{-1}
\end{equation}
(where  $\alpha_Y,\alpha_L$ and $\alpha_c$ are the $U(1)_Y,SU(2)_L$
and $SU(3)_c$ couplings respectively) is a function of only $M_R$ and $M_c$.

In what follows, we perform a one-loop analysis, and in order to
estimate $M_c$ and $M_R$ we can safely ignore the difference between
$M_W$  and $\Lambda_{SUSY}$ noting that it becomes relevant at the
two-loop level. It is then easy to show that

\begin{description}

\item[{\bf a)}] for $M_R^2 \geq M_W M_c$, 
\begin{equation}
2\pi \Delta\alpha^{-1} = 12 \ln\frac{M_c}{M_W} + 8  \ln\frac{M_R}{M_W}
,\label{chaina}
\end{equation}
and
\item[{\bf b)}]for $M_R^2 \leq M_W M_c$,
\begin{equation}
2\pi \Delta\alpha^{-1} = 14 \ln\frac{M_c}{M_W} + 4  \ln\frac{M_R}{M_W}
.
\label{chainb}
\end{equation}
\end{description}
In the MSSM, one would get
\begin{equation}
2\pi \Delta\alpha^{-1} = 12 \ln\frac{M_U}{M_W}
\end{equation}
where $M_U$ is the unification scale, $M_U \simeq 10^{16} GeV$. 

It is easy to see that in this approximation
\begin{equation}
10^{10} GeV \leq M_c \leq 10^{14} GeV
\label{rangemc}
\end{equation}
while at the same time
\begin{equation}
10^{10} GeV \geq M_R \geq 10^{3} GeV
\label{rangemr}
\end{equation}

As it is clear from (\ref{chaina}) and (\ref{chainb}), lower $M_R$
implies bigger $M_c$. It is remarkable that unification constraints
allow for the parity breaking scale to be experimentally
detectable. Of course since the see-saw mechanism requires $m_\nu
\propto M_R^{-1}$, small neutrino masses prefer larger $M_R$. 

A comment is noteworthy here. This is true if we assume that $m_D$
(the Dirac neutrino mass) is of the order of the charged lepton or
quark masses. In the minimal theory one is tempted to use only a
(2,2,1) Higgs to break $SU(2)_L\times U(1)_Y$. Then one has 
\begin{equation}
m_D = m_u , \quad m_\ell = m_d
\label{badly}
\end{equation}
for the down quarks ($d$), charged leptons ($\ell$), up quarks ($u$)
 and neutrino Dirac ($D$)  masses. But $m_\ell = m_d$ (at $M_c$)
fails badly for the first two generations and so (\ref{badly}) cannot
really be trusted. One can add a (2,2,15) in order  to correct
(\ref{badly}), in which case the mass spectrum looks like 
\begin{eqnarray}
m_u & = & y^1 v_u^1 + y^{15} v_u^{15}\nonumber \\
m_D & = & y^1 v_u^1 - 3 y^{15} v_u^{15} \nonumber \\
m_d & = & y^1 v_d^1 + y^{15} v_d^{15} \nonumber \\
m_\ell & = & y^1 v_d^1 - 3 y^{15} v_d^{15}
\label{goodly}
\end{eqnarray}

Clearly, the theory can be made realistic without any change in the
unification predictions since only one bi-doublet remains light after
the usual fine-tuning. The price is arbitrariness at $m_D$ and the
inability to predict neutrino masses.
Low $M_R$ is then  phenomenologically
allowed and in accord with $SU(4)_c$ unification.

A positive note.  As is well-known, in LR theories the see-saw
mechanism is {\em not}  type I, i.e. it contains an additional piece
$
\propto <\Delta> \simeq M_W^2/M_R$ \cite{mw80,ms81}. What happens is the following. The
symmetry allows for a coupling in the potential 
\begin{equation}
\Delta V = \lambda \Delta \phi^2 \Delta^c + M^2 \Delta^2
\label{deltaphidelta}
\end{equation}
which gives a small vev to $\Delta$
\begin{equation}
<\Delta > = \lambda \frac{<\phi>^2 M_R}{M^2} \simeq \lambda \frac{M_W^2}{M_R}
\end{equation}
Now, in supersymmetry this does not happen, at least at the
renormalizable level. We can have higher-dimensional terms in the
superpotential
\begin{equation}
\Delta W = \frac{1}{M_{Pl}} \Delta \phi^2 \Delta^c
\label{nonren}
\end{equation}
which imply a tiny, negligible vev $<\Delta> \propto M_W^2/M_{Pl}$.

One could also generate such terms if there are interactions
\begin{equation}
W= \phi^2 S + \Delta \Delta^c S + M S^2
\label{ws}
\end{equation}
where $S = (3,3,1)$ under $G_{PS}$. Integrating out $S$ would then
give (\ref{nonren}) with $M_{Pl} \to M$. 

Another possibility would be 
\begin{equation}
W = \phi\Delta X + \phi \Delta^c \bar X + M X \bar X
\label{wx}
\end{equation}
where $X= 2,2,\bar 10)$ and $\bar X = 2,2,10)$ under $G_{PS}$. The
absence of the $S, X, \bar X$ fields guarantees a type I see-saw at
the supersymmetric level.

Now, once supersymmetry is broken, one can generate a nonvanishing but
negligible vev for $\Delta$ \cite{amrs99}:
\begin{equation}
<\Delta > \simeq \left(\frac{m_{3/2}}{M_c} \right)^2 \frac{m_D^2}{M_R}
\end{equation}

which contributes  by a tiny factor $(m_{3/2}/M_c)^2\leq
10^{-14}$ to the usual see-saw mass term $m_\nu \simeq
m_D^2/m_{\nu_R}$.

In short, the see-saw mechanism takes its canonical form in the
minimal model. Of course, since we do not know $m_D$ and the
right-handed neutrino masses we cannot predict neutrino masses
precisely, but the type I form serves for the leptogenesis
scenario. Namely, in this case it is the Dirac Yukawa couplings which
are responsible for bringing the right-handed neutrinos into
equilibrium and also for their decay. Thus one can set constraints on
$m_D$ and right-handed neutrino masses.

\section{Effects of non-renormalizable terms}

   The crucial ingredient in obtaining (\ref{rangemc}) and
(\ref{rangemr}) is the existence of doubly charged supermultiplets at
the scale $\Lambda_{SUSY} \simeq TeV$. We have not considered
  higher
dimensional operators of the form $(\Delta^c \bar\Delta^c)^2/M_{Pl}$,
which are likely to be present and would
cure the problem of potential breaking of the electromagnetic
charge \cite{mr96}, even if the SUSY breaking terms preserve $SU(2)_R$.
These terms would also give a mass to the doubly charged states,
relevant for $M_R^2 \geq M_W M_{Pl}$. The unification constraint is now
\begin{equation}
2 \pi \Delta\alpha^{-1} = 12 \ln\frac{M_c}{M_W} -8  \ln\frac{M_r}{M_W}
+ 8  \ln\frac{M_{Pl}}{M_W}
\end{equation}
The only consistent solution is the single-step breaking
\begin{equation}
M_R\simeq M_c\simeq 10^{10}
GeV
\end{equation}
which guarantees the lightness of the doubly-charged
supermultiplets. Since the color octets however become heavy, it is
hard to distinguish this theory from the LR models. The only hope
would be to find the relatively light magnetic monopoles, but it is
hard to imagine an effective production mechanism in the context of
supersymmetry.

Now, could there be another source of these masses? In principle,
yes. Namely, the large representations we need to achieve the
symmetry breaking imply the loss of asymptotic freedom above
$M_c$. In fact, the gauge coupling becomes strong at the scale $M_F
\simeq 10 M_c$. Thus, strictly speaking we could imagine operators of
the type $(\Delta^c \bar\Delta^c)^2/M_{F}$ which give a mass
$M_R^2/10 M_c$ to the doubly-charged states. With the color octets and
doubly charged particles having basically the same mass, the
unification constraint turns out to be
\begin{equation}
2 \pi \Delta\alpha^{-1} = 20 \ln\frac{M_c}{M_W} - 8  \ln\frac{M_r}{M_W}
\end{equation}
Keeping in mind that these effects are important for $M_R^2 \geq M_c
M_W$, one gets
\begin{equation}
10^{12} GeV \leq M_c \leq 10^{16} GeV \; ; \quad 10^{7} GeV \leq M_R 
\leq 10^{16} GeV
\end{equation}

In other words, for $M_R \leq 10^7 GeV$, the colour octets and doubly
charged states lie at the $TeV$ scale, since the non-renormalizable
terms play no role whatsoever. For $M_R $ bigger than $10^7 GeV$, it
becomes harder to find these states, but they still remain comparable in mass.
 This could be the smoking gun of 
supersymmetric
Pati-Salam theory.

\section{Summary and outlook}

The minimal renormalizable supersymmetric Pati-Salam theory offers the
exciting possibility of low scale parity restoration, even as low as
$TeV$, thus making it accessible to experiment. The important, crucial
prediction of the theory is the existence of doubly charged  and
color octet supermultiplets with an almost degenerate mass. The
discovery of these states would be a  signal of four colour
unification.  

  As discussed
in the paper,  for $10^3 GeV \leq M_R \leq 10^7 GeV$ ($ 10^{12} GeV
\leq M_c \leq 10^{14} GeV $), these particles could be discovered by LHC
 at the
$TeV$ scale. In all honesty, such a low $M_R$ requires some
fine-tuning in order to achieve a small neutrino mass. For larger
$M_R$ these particles become heavier and less accessible to experiment,
however non-renormalizable effects suppressed by a fundamental scale
around $10 M_c$ still guarantee that they have comparable masses.
 
Breaking of PS symmetry
  at $M_c$ implies the existence of $U(1)_{B-L}$ monopoles.
 with mass $m_M \simeq 10 M_c$. If produced in a phase transition via
 the Kibble mechanism, the requirement that
their density be less than the critical density then implies
$M_c \leq 10^{12} GeV$.
In other words, if non-renormalizable terms cut off by $M_F \sim 10 M_c$ 
are present,  we have the usual
GUT (superheavy) monopole and one can invoke inflation or some other
mechanism in order to get rid of them. Unfortunately, in this case the
number of monopoles, if not zero, is not predictable at all. The 
single-step breaking at $M_c\sim M_R \sim 10^{10} GeV$, on the other hand,
 offers the interesting possibility of potentially detectable intermediate
 mass monopoles. This however is a delicate point, since as is usual
in supersymmetric theories, high-temperature effects would lead to a
false vacuum problem. 
Namely, the phase transition may not occur at
all, rending the theory unrealistic
 (for recent work and references see \cite{bggs01}).
 Whether if this can be avoided
and an estimate of the monopole density can be made is beyond the
scope of this paper. 

\acknowledgments
We thank Borut Bajc for discussions. The work of A.M. was partially supported
by CDCHT-ULA (Project C-1073-01-05-A). The work of G.S. is partially 
supported by EEC (TMR contracts ERBFMRX-CT960090
 and HPRN-CT-2000-00152). 
A.M. wishes to thank ICTP for hospitality during the course of this work.

\end{document}